# Ferromagnetic and antiferromagnetic dimer splittings in $LaMn_{0.1}Ga_{0.9}O_3$

A. Furrer,[1] E. Pomjakushina,[2] V. Pomjakushin,[1] J. P. Embs,[1] and Th. Strässle[1]

[1] Laboratory for Neutron Scattering, Paul Scherrer Institut, CH-5232 Villigen PSI, Switzerland

[2] Laboratory for Developments and Methods, Paul Scherrer Institut, CH-5232 Villigen PSI, Switzerland

**Abstract:**

Inelastic neutron scattering was employed to study the magnetic excitations of $Mn^{3+}$ dimers in $LaMn_{0.1}Ga_{0.9}O_3$. The nearest-neighbor interaction of $Mn^{3+}$ ions is ferromagnetic in the basal (a,b)-plane, but antiferromagnetic along the c-direction, thus two different types of dimer excitations are simultaneously present in the experiments. From the observed energy spectra we derive Heisenberg-type exchange interactions $J_{ab}$=0.210(4) meV and $J_c$=-0.285(5) meV as well as an axial anisotropy parameter D=0.036(6) meV. These parameters considerably differ from those derived for the isostructural parent compound $LaMnO_3$ due to structural effects.

# I. INTRODUCTION

The rich phase diagrams of hole-doped manganese oxides of perovskite type, $R_{1-x}A_xMnO_3$, where R is a rare-earth element (La, Pr, Nd, …) and A is a divalent element (Ca, Sr, Ba, …), were established roughly half a century ago [1]. These compounds received renewed attention due to the discovery of the phenomenon of giant magnetoresistance upon the replacement of $La^{3+}$ ions by $A^{2+}$ ions [2]. The parent compound $LaMnO_3$ is an antiferromagnetic insulator in which an orbital ordering is established due to the cooperative Jahn-Teller effect breaking the degeneracy of the electronic configuration of the $Mn^{3+}$ ions. This particular orbital ordering is responsible for the A-type magnetic structure determined by Wollan and Koehler [3], with a ferromagnetic coupling of the $Mn^{3+}$ spins in the basal (a,b)-plane and an antiferromagnetic coupling in the c-direction perpendicular to this plane. The doping with $A^{2+}$ ions promotes part of the trivalent manganese ions into the tetravalent state, and a ferromagnetic ground state is realized due to the double exchange mechanism between $Mn^{3+}$ and $Mn^{4+}$ [4]. $LaMnO_{3+\delta}$ compounds with oxygen excess δ exhibit a ferromagnetic ground state as well [5]. Surprisingly, the partial replacement of $Mn^{3+}$ ions by non-magnetic trivalent ions like $Ga^{3+}$ also induces ferromagnetism [6], but in this case the double exchange mechanism cannot be applied. Therefore a detailed study of the exchange interactions in the compound $LaMn_xGa_{1-x}O_3$ appears to be appropriate. Moreover, it has been suggested that biquadratic exchange interactions are important to understand the complex phase diagrams of the manganese perovskites, particularly those involving the heavy rare earths [7]. Inelastic neutron scattering experiments of isolated clusters of magnetic ions are ideally suited to address these questions [8], hence, we present here a neutron scattering study of the exchange and anisotropy parameters in the magnetically diluted compound $LaMn_{0.1}Ga_{0.9}O_3$.

The compound LaMn$_{0.1}$Ga$_{0.9}$O$_3$ is isostructural to LaMnO$_3$ (orthorhombic space group *Pbnm*). The partial substitution of Mn$^{3+}$ ions by non-magnetic Ga$^{3+}$ ions results in the creation of Mn$^{3+}$ monomers, dimers, trimers, etc. The deviation of the magnetic susceptibility from the Curie-Weiss law at low temperatures is a fingerprint of the existence of Mn$^{3+}$ multimers [9]. The formation of Mn$^{3+}$ dimers occurs with a statistical maximum for the chosen 90% dilution with Ga$^{3+}$ ions. Due to the nature of the nearest-neighbor exchange coupling, ferromagnetic in the basal (a,b)-plane and antiferromagnetic along the c-direction, the number of ferromagnetically coupled Mn$^{3+}$ dimers is twice as large as that of antiferromagnetically coupled Mn$^{3+}$ dimers. We therefore encounter the interesting situation of the simultaneous presence of two different types of dimer exchange splitting patterns, which is a new and challenging aspect in the experimental study of magnetic cluster systems.

The present work is organized as follows. The sample synthesis and structural characterization as well as the experimental procedures are described in Sec. II, followed in Sec. III by a summary of the spin Hamiltonian and neutron cross-section for spin dimers. The experimental results and their analyses are presented in Sec. IV. Finally, a brief discussion and some conclusions are given in Sec. V.

## II. EXPERIMENTAL

### A. Sample Synthesis

The sample of LaGa$_{0.9}$Mn$_{0.1}$O$_3$ was synthesised by a solid state reaction using La$_2$O$_3$, Ga$_2$O$_3$ and MnO$_2$ of a minimum purity of 99.99%. The respective amounts of starting reagents were mixed, milled and calcinated at 1250°C during 24 h in air. Then the resulting powders were milled, pressed into pellets and sintered in air at 1250ºC during 50 h. Finally the sample was milled and annealed in an argon atmosphere at 1350ºC during 40 h to remove possible overstoichiometric oxygen. The phase purity was checked with x-ray powder diffraction using a D8 Advance Bruker AXS diffractometer with Cu K$\alpha$ radiation.

**B. Neutron Diffraction**

The neutron powder diffraction experiments were carried out at the spallation neutron source SINQ at PSI Villigen using the high-resolution diffractometer for thermal neutrons HRPT [10] ($\lambda$=1.494 Å, high intensity mode with $\delta d/d=1.8\cdot 10^{-3}$) at the temperatures 300 K and 1.5 K. The refinements of the crystal structures were carried out with the program FULLPROF [11], with use of its internal tables for scattering lengths. The resulting structural data are listed in Tables I and II. Our results are in good agreement with those published by Blasco et al. [9]. The refined numbers of the stoichiometries for Mn and O confirm the chemical composition of the title compound.

**C. Neutron Spectroscopy**

The inelastic neutron scattering experiments were carried out with use of the high-resolution time-of-flight spectrometer FOCUS [13] at the spallation neutron source SINQ at PSI Villigen. The measurements were performed with incoming neutron energies $E_i$=7.99, 4.23, and 2.47 meV in the time-focusing mode, with instrumental energy resolutions at the elastic position of 410, 200, and 54 µeV, respectively. The scattered neutrons were detected by an array of $^3$He counters covering a large range of scattering angles 10º$\leq\Phi\leq$130º. The sample was enclosed in an aluminum cylinder (12 mm diameter, 45 mm height) and placed into a He cryostat to achieve temperatures 1.5$\leq$T$\leq$75 K. Additional experiments were performed for the empty container as well as for vanadium to allow the correction of the raw data with respect to background, detector efficiency, absorption, and detailed balance according to standard procedures.

## III. THEORETICAL BACKGROUND

We base the analysis of the $Mn^{3+}$ dimer transitions on the spin Hamiltonian

$$H = -2J\mathbf{s_1}\cdot\mathbf{s_2} - D\left[\left(s_1^z\right)^2 + \left(s_2^z\right)^2\right] \quad (1)$$

where $\mathbf{s_i}$ denotes the spin operator of the magnetic ions, J the bilinear exchange interaction, and D the axial single-ion anisotropy parameter. H commutes with the total spin $\mathbf{S}=\mathbf{s_1}+\mathbf{s_2}$, thus S is a good quantum number to describe the spin states as |S,M> with -S≤M≤S. For D=0 and identical magnetic ions ($s_1=s_2$) the eigenvalues of Eq. (1) are degenerate with respect to the quantum number M:

$$E(S) = -J[S(S+1) - 2s_i(s_i+1)] \quad (2)$$

The energy level sequence follows the well-known Landé interval rule:

$$E(S) - E(S-1) = -2JS \quad (3)$$

For $Mn^{3+}$ ions with $s_i=2$, ferromagnetic (J>0) and antiferromagnetic (J<0) exchange give rise to a nonet (S=4) and a singlet (S=0) ground state, respectively, as illustrated in Fig. 1. A non-zero anisotropy term (D≠0) has the effect of splitting the spin states |S> into the states |S,±M>. For D<0 the energetic ordering of the sublevels |S,±M> has to be reversed in Fig. 1.

For spin dimers the neutron cross-section for a transition from the initial state |S⟩ to the final state |S'⟩ is defined by [14]

$$\frac{d^2\sigma}{d\Omega d\omega} = \frac{N}{Z}(\gamma r_0)^2 \frac{k'}{k} F^2(Q) \exp\{-2W(Q)\} \exp\left\{-\frac{E(S)}{k_B T}\right\}$$
$$\times \frac{4}{3}\left[1 + (-1)^{\Delta S} \frac{\sin(QR)}{QR}\right] |T_1|^2 \delta\{\hbar\omega + E(S) - E(S')\} \quad (4)$$

where N is the total number of spin dimers in the sample, Z the partition function, k and k' the wave numbers of the incoming and scattered neutrons, respectively, Q the modulus of the scattering vector **Q=k-k'**, F(Q) the magnetic form factor, exp{-2W(Q)} the Debye-Waller factor, R the distance between the two dimer spins, $T_i = \langle S'\|T_i\|S\rangle$ ($T_1 = T_2$) the reduced transition matrix element defined in Ref. 14, and $\hbar\omega$ the energy transfer. The remaining symbols have their usual meaning. The transition matrix element carries essential information to derive the selection rules for spin dimers:

$$\Delta S = S - S' = 0, \pm 1 \; ; \; \Delta M = M - M' = 0, \pm 1 \; . \quad (5)$$

The transitions observed in the present work for ferromagnetically and antiferromagnetically coupled $Mn^{3+}$ dimers are marked in Fig. 1 by the arrows A and B, respectively.

## IV. RESULTS AND DATA ANALYSIS

Energy spectra of neutrons scattered from $LaMn_{0.1}Ga_{0.9}O_3$ with $E_i$=7.99 meV are shown for different temperatures in Fig. 2. The spectrum at T=1.5 K is characterized by a partly resolved line at an energy transfer of 1.7 meV. No additional lines show up with increasing temperature at higher energy transfers. According to the energy splitting pattern of Fig. 1, the transition with largest energy is the |4>→|3> transition associated with ferromagnetically coupled $Mn^{3+}$ dimers. We therefore attribute the line at 1.7 meV to this transition, which is completely resolved in the spectra taken with improved energy resolution ($E_i$=4.23 meV) shown in Fig. 3. In addition, the T=1.5 K data exhibit a shoulder at an energy transfer of 0.5 meV, which could be resolved with further improved energy resolution ($E_i$=2.47 meV) as illustrated in Fig. 4. As any line appearing at low temperatures (T=1.5 K) has to be a transition out of the ground state, we attribute the line at 0.5 meV to the |0>→|1> transition associated with antiferromagnetically coupled $Mn^{3+}$ dimers. The intrinsic linewidths of both peaks A and B are considerably larger than the instrumental energy resolution because of the splitting of the spin states |S> into the substates |S,±M> resulting from the single-ion anisotropy as visualized in Fig. 1. A further contribution to the intrinsic linewidth can be attributed to the different local geometries around the $Mn^{3+}$ ions.

We start the data analysis with Fig. 4. The data taken at T=1.5 K were fitted by two Gaussian functions with equal linewidth, and the background was approximated by an exponential function. The two Gaussians are centered at energy transfers of 0.52(2) meV (peak $B_1$) and 0.68(3) meV (peak $B_2$) with an intensity ratio $I(B_1)/I(B_2)$=2.2(5). We can readily identify the peaks $B_1$ and $B_2$ by the |0,0>→|1,±1> and the |0,0>→|1,0> dimer transitions, respectively, since the calculated intensity ratio is $I(B_1)/I(B_2)$=2.4. Application of the Hamiltonian (4) then yields the parameters $J_c$=-0.285(5) meV and D=0.036(6) meV. With increasing temperature the energy spectra of Fig. 4 are completely smeared out due to excited-state transitions.

We now turn to the data of Fig. 3. The peak A, corresponding to the |4>→|3> transition associated with ferromagnetically coupled $Mn^{3+}$ dimers, is actually a superposition of eleven individual transitions of type

|4,±M⟩→|3,±M'⟩ as illustrated in Fig. 1, which could not be resolved in the experiments. For the data analysis we therefore kept the anisotropy parameter fixed at the value D=0.036 meV determined above and described the energy spectra by eleven Gaussian functions with equal linewidth (and assuming a linear background). The spectral strengths of the eleven Gaussians were fixed at the calculated probabilities associated with the particular |4,±M⟩→|3,±M'⟩ transitions. The least-squares fitting procedure converged to the parameter $J_{ab}$=0.210(4) meV. The resulting energy spectra described by lines in Fig. 3 are in excellent agreement with the experimental data; in particular, the decrease of both the integrated intensities and the mean energy transfers with increasing temperature are nicely reproduced. Our interpretation is also supported by the Q dependence of the peak intensities as shown in Fig. 5, which are in agreement with the prediction of the cross-section formula (4). The final model parameters are listed in Table III.

## V. DISCUSSION AND CONCLUSIONS

It is interesting to compare the model parameters obtained for $LaMn_{0.1}Ga_{0.9}O_3$ with those of the isostructural parent compound $LaMnO_3$. Hirota et al. [15] and Moussa et al. [16] measured the spin-wave dispersion of $LaMnO_3$ by inelastic neutron scattering. Their data analysis was based on a Heisenberg Hamiltonian including an anisotropic interaction:

$$H = -2\sum_{i>j} J_{ij}\, \mathbf{s_i} \cdot \mathbf{s_j} - H_{aniso} \qquad (6)$$

Hirota et al. [15] described the anisotropy term by

$$H_{aniso} = -g\mu_B H_A \left( \sum_m s_m^z - \sum_n s_n^z \right) \tag{7}$$

where $H_A$ is an effective anisotropy field, and the summation indices m and n run over the up and down spins, respectively. Moussa et al. [16] described the anisotropy by the axial single-ion Hamiltonian

$$H_{aniso} = -D\sum_i \left(s_i^z\right)^2 \tag{8}$$

Using the Holstein-Primakoff approximation [17] up to second order, an explicit formula for the spin-wave dispersion is obtained from the Hamiltonian (6) [15,16], in which the anisotropy parameters of Eqs (7) and (8) are related through $D=g\mu_B H_A/4$. Only two exchange integrals were necessary to fit the observed spin-wave dispersion: $J_{ab}$ describing the ferromagnetic coupling between nearest $Mn^{3+}$ neighbors in the basal (a,b)-plane and $J_c$ between nearest neighbors along the c-direction. The model parameters are listed in Table III.

There are marked differences between the model parameters for x=1 [15,16] and x=0.1 (present work). These differences are largely due to structural effects. The replacement of $Mn^{3+}$ by $Ga^{3+}$ results in a continuous decrease of the tetragonal distortion of the $MnO_6$ octahedron [9], which is practically regular for x≤0.4. This explains the drastic decrease of the anisotropy parameter D when going from x=1 to x=0.1. On the other hand, the exchange parameter $J_c$ appears to be unaffected by the doping with Ga, whereas $J_{ab}$ is reduced by a factor of 2.

The exchange parameters $J_{ab}$ and $J_c$ are mediated through the Mn-O2-Mn and Mn-O1-Mn bridges, respectively. The size and the sign of the exchange parameters may depend on the bond angles φ which according to Table II slightly increase when going from x=1 to x=0.1. The exchange coupling $J_c$ corresponds to π-bonding which is weakly dependent on the bond angle, so that the value of $J_c$ remains almost unaffected by the dilution with Ga. On the other hand, the exchange coupling $J_{ab}$ is governed by σ-bonding, for which its size is expected to scale according to cos(φ) [18]. We therefore expect an increase of $J_{ab}$ upon dilution with Ga, however, our experiments do not support this picture. Obviously the local geometry around the Mn ions must be different from the averaged geometry which is largely dictated by the Jahn-Teller non-active Ga ions. Evidence for local structures was obtained e.g. in the related compounds $La_{1-x}Ca_xMnO_3$ by using the atomic pair-distribution-function analysis of neutron powder-diffraction data [19]. The weakening of $J_{ab}$ upon dilution with Ga is presumably due to a weakening of the cooperative Jahn-Teller effect associated with the minority $Mn^{3+}$ ions that causes the ferromagnetic nature of the in-plane exchange.

In conclusion, we were able to derive the relevant exchange and anisotropy parameters of the manganese compound $LaMn_{0.1}Ga_{0.9}O_3$ directly from the ground-state transitions observed for both ferromagnetically and antiferromagnetically coupled $Mn^{3+}$ dimers. The analysis of the observed energy spectra had to be restricted to the T=1.5 K data, since already a slight enhancement of the temperature completely smears out the magnetic response for energy transfers less than 1 meV (see Fig. 4). This is due to scattering contributions from excited-state transitions which become increasingly populated at higher temperatures. It was therefore impossible to derive any information about the possible existence of biquadratic interactions [7] which requires the observation of well resolved excited-state transitions [20]. Nevertheless, our results contribute to the understanding of the observed superparamagnetism in $LaMn_xGa_{1-x}O_3$ (x≤0.2) [9] in terms of ferromagnetic and antiferromagnetic $Mn^{3+}$ dimer excitations which - although simultaneously present in the experiments – could be successfully disentangled from each other.

**ACKNOWLEDGMENTS**

This work was performed at the Swiss Spallation Neutron Source SINQ, Paul Scherrer Institut (PSI), Villigen, Switzerland. Financial support by the Swiss National Science Foundation through the NCCR MaNEP project is gratefully acknowledged.

**References**

[1]   J. B. Goodenough, Phys. Rev. **100**, 564 (1955).

[2]   R. von Helmolt, J. Wecker, B. Holzapfel, L. Schultz, and K. Samwer, Phys. Rev. Lett. **71**, 2331 (1993).

[3]   E. O. Wollan and W. C. Koehler, Phys. Rev. **100**, 545 (1955).

[4]   C. Zener, Phys. Rev. **82**, 403 (1951).

[5]   C. Ritter, J. M. de Teresa, P.A. Algarabel, C. Marquina, J. Blasco, J. Garcìa, S. Oseroff, and S.-W. Cheong, Phys. Rev. B **56**, 8902 (1997).

[6]   J.-S. Zhou, H. Q. Yin, and J. B. Goodenough, Phys. Rev. B **63**, 184423 (2001).

[7]   T. A. Kaplan, Phys. Rev. B **80**, 012407 (2009).

[8]   A. Furrer, Int. J. Mod. Phys. B **24**, 3653 (2010).

[9]   J. Blasco, J. Garcia, J. Campo, M. C. Sanchez, and G Subias, Phys. Rev. B **66**, 174431 (2002).

[10] P. Fischer, G. Frey, M. Koch, M. Koennecke, V. Pomjakushin, J. Schefer, R. Thut, N. Schlumpf, R. Bürge, U. Greuter, S. Bondt. and E. Berruyer, Physica B **276-278**, 146 (2000).

[11] J. Rodriguez-Carvajal, Physica B **192**, 55 (1993).

[12] J. Rodriguez-Carvajal, M. Hennion, F. Moussa, A. Moudden, L. Pinsard, and A. Revcolevschi, Phys. Rev. B **57**, 3189 (1998).

[13] J. Mesot, S. Janssen, L. Holitzner, and R. Hempelmann, J. Neutron Res. **3**, 293 (1996).

[14] A. Furrer and H. U. Güdel, J. Magn. Magn. Mater. **14**, 256 (1979).

[15] K. Hirota, N. Kaneko, A. Nishizawa, and Y. Endoh, J. Phys. Soc. Japan **65**, 3736 (1996).

[16] F. Moussa, M. Hennion, J. Rodriguez-Carvajal, H. Moudden, L. Pinsard, and A. Revcolevschi, Phys. Rev. B **54**, 15149 (1996).

[17] T. Holstein and H. Primakoff, Phys. Rev. **58**, 1094 (1940).

[18] J. B. Goodenough, *Magnetism and Chemical Bond* (Interscience Publishers, New York, 1963).

[19] S. J. L. Billinge, R. G. Di Francesco, G. H. Kwei, J. J. Neumeier, and J. D. Thompson, Phys. Rev. Lett. **77**, 715 (1996).

[20] Th. Strässle, F. Juranyi, M. Schneider, S. Janssen, A. Furrer, K. W. Krämer, and H. U. Güdel, Phys. Rev. Lett. **92**, 257202 (2004).

TABLE I. Crystal structure parameters and reliability factors for $LaMn_{0.1}Ga_{0.9}O_3$ in the structure model *Pbnm* (No. 62) at temperatures 1.5 K and 300 K. La is in the (4c)-position (x,y,1/4), Ga/Mn in (4a) (0,0,0), O1 in (4c), and O2 in the general (8d)-position. p denotes the stoichiometry (for Mn and O).

| | | T=1.5 K | T=300 K |
|---|---|---|---|
| a [Å] | | 5.51251(2) | 5.52498(2) |
| b [Å] | | 5.48691(2) | 5.49518(2) |
| c [Å] | | 7.78929(4) | 7.79032(3) |
| La: | x | 0.0051(2) | 0.0039(2) |
| | y | 0.5214(1) | 0.5185(1) |
| O1: | x | -0.0687(3) | -0.0667(2) |
| | y | -0.0087(2) | -0.0077(2) |
| O2: | x | 0.2265(1) | 0.2285(1) |
| | y | 0.2745(1) | 0.2725(1) |
| | z | 0.0367(1) | 0.03645(8) |
| La: | B [Å$^2$] | 0.08(1) | 0.44(1) |
| Ga/Mn: | B [Å$^2$] | 0.093(14) | 0.28(1) |
| O1: | B [Å$^2$] | 0.384(17) | 0.63(2) |
| O2: | B [Å$^2$] | 0.28(1) | 0.57(1) |
| Mn: | p | 0.106(2) | 0.105(1) |
| O: | p | 3.030(8) | 3.008(7) |
| $R_p$ [%] | | 2.88 | 2.22 |
| $R_{wp}$ [%] | | 3.70 | 2.90 |
| $R_{exp}$ [%] | | 2.21 | 1.86 |
| $\chi^2$ | | 2.80 | 2.43 |

TABLE II. Bond lengths d and bond angles φ around $Mn^{3+}$ ions for $LaMn_{0.1}Ga_{0.9}O_3$ at T=1.5 K and 300 K and for $LaMnO_3$ [a=5.5367(1), b=5.7473(1), c=7.6929(2)] at room temperature [12].

|  | $LaMn_{0.1}Ga_{0.9}O_3$ |  | $LaMnO_3$ |
| --- | --- | --- | --- |
| T [K] | 1.5 | 300 | 300 |
| d(Mn-O1) [Å] | 1.9844(3) | 1.9826(2) | 1.9680(3) |
| d(Mn-O21) [Å] | 1.9715(7) | 1.9792(7) | 1.907(1) |
| d(Mn-O22) [Å] | 1.9768(7) | 1.9732(7) | 2.178(1) |
| φ(Mn-O1-Mn) [°] | 157.819(13) | 158.435(8) | 155.48(2) |
| φ(Mn-O2-Mn) [°] | 160.10(3) | 160.67(3) | 155.11(5) |

TABLE III. Exchange and anisotropy parameters derived for $LaMn_xGa_{1-x}O_3$. $J_{ab}$ denotes the ferromagnetic coupling between nearest $Mn^{3+}$ neighbors in the basal (a,b)-plane, and $J_c$ the antiferromagnetic coupling between nearest neighbors along the c-direction. The exchange parameters for x=1 correspond to half the values given in Refs 15 and 16 due to the prefactor 2 introduced in Eq. (6).

| x | T [K] | $J_{ab}$ [meV] | $J_c$ [meV] | D [meV] | Reference |
|---|---|---|---|---|---|
| 1.0 | ≈10 | 0.418(6) | -0.302(14) | 0.152(28) | [15] |
| 1.0 | 20 | 0.42(3) | -0.29(2) | 0.165(9) | [16] |
| 0.1 | 1.5 | 0.210(4) | -0.285(5) | 0.036(6) | present work |

**Figure Captions**

FIG. 1. Schematic sketch of energy level splittings of magnetic dimers with $s_i=2$. The arrows mark the transitions observed for $LaMn_{0.1}Ga_{0.9}O_3$ in the present work. The energy splittings of the states $|S\rangle$ into the substates $|S,\pm M\rangle$ resulting from the single-ion anisotropy are enhanced for better visualization.

FIG. 2. (Color online) Energy spectra of neutrons scattered from $LaMn_{0.1}Ga_{0.9}O_3$ at different temperatures. The incoming neutron energy was 7.99 meV.

FIG. 3. (Color online) Energy spectra of neutrons scattered from $LaMn_{0.1}Ga_{0.9}O_3$ at different temperatures. The incoming neutron energy was 4.23 meV. The lines are the result of a least-squares fitting procedure as explained in the text.

FIG. 4. (Color online) Energy spectra of neutrons scattered from $LaMn_{0.1}Ga_{0.9}O_3$ at different temperatures. The incoming neutron energy was 2.47 meV. The lines are the result of a least-squares fitting procedure as explained in the text.

FIG. 5. (Color online) Q dependence of the neutron cross-section associated with transitions of ferromagnetically coupled $Mn^{3+}$ dimers in $LaMn_{0.1}Ga_{0.9}O_3$. The line corresponds to Eq. (4).

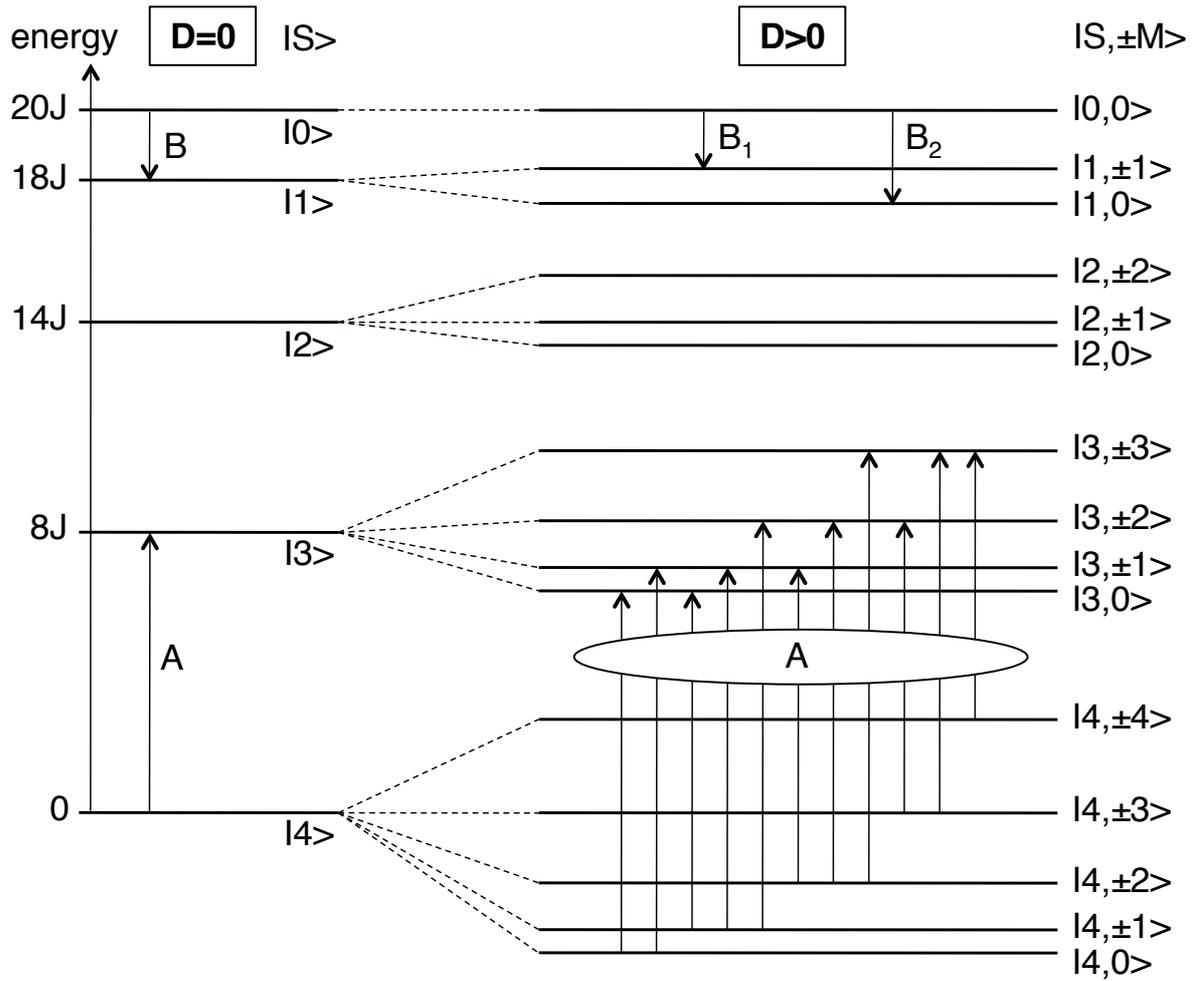

FIG. 1. Schematic sketch of energy level splittings of magnetic dimers with $s_i=2$. The arrows mark the transitions observed for $LaMn_{0.1}Ga_{0.9}O_3$ in the present work. The energy splittings of the states $|S\rangle$ into the substates $|S,\pm M\rangle$ resulting from the single-ion anisotropy are enhanced for better visualization.

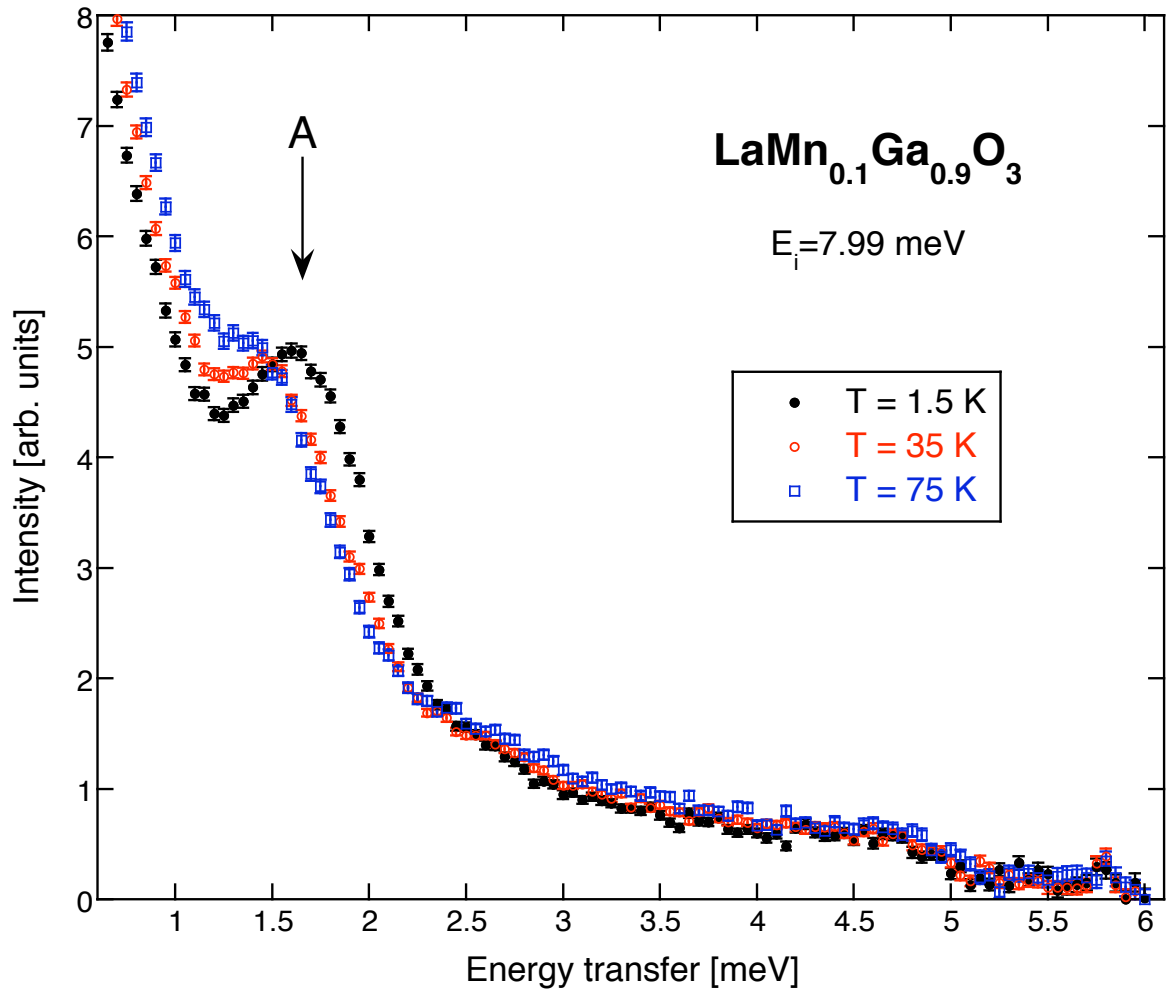

FIG. 2. (Color online) Energy spectra of neutrons scattered from $LaMn_{0.1}Ga_{0.9}O_3$ at different temperatures. The incoming neutron energy was 7.99 meV.

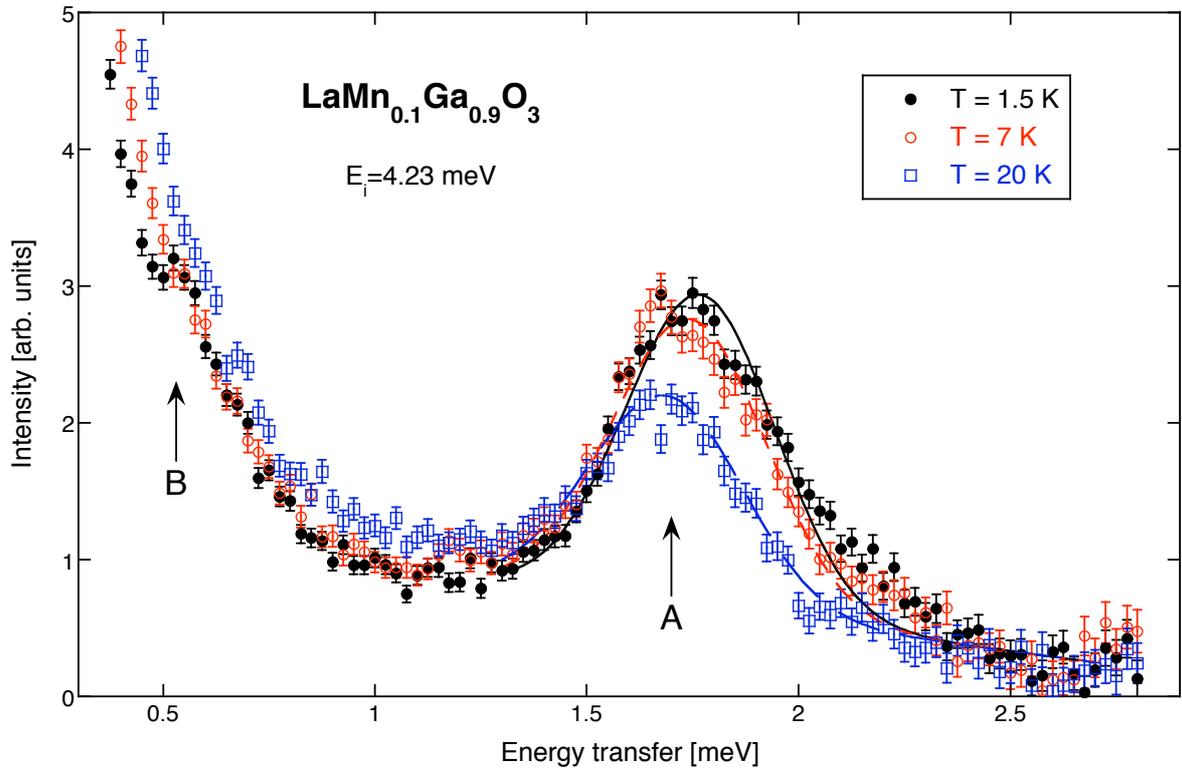

FIG. 3. (Color online) Energy spectra of neutrons scattered from $LaMn_{0.1}Ga_{0.9}O_3$ at different temperatures. The incoming neutron energy was 4.23 meV. The lines are the result of a least-squares fitting procedure as explained in the text.

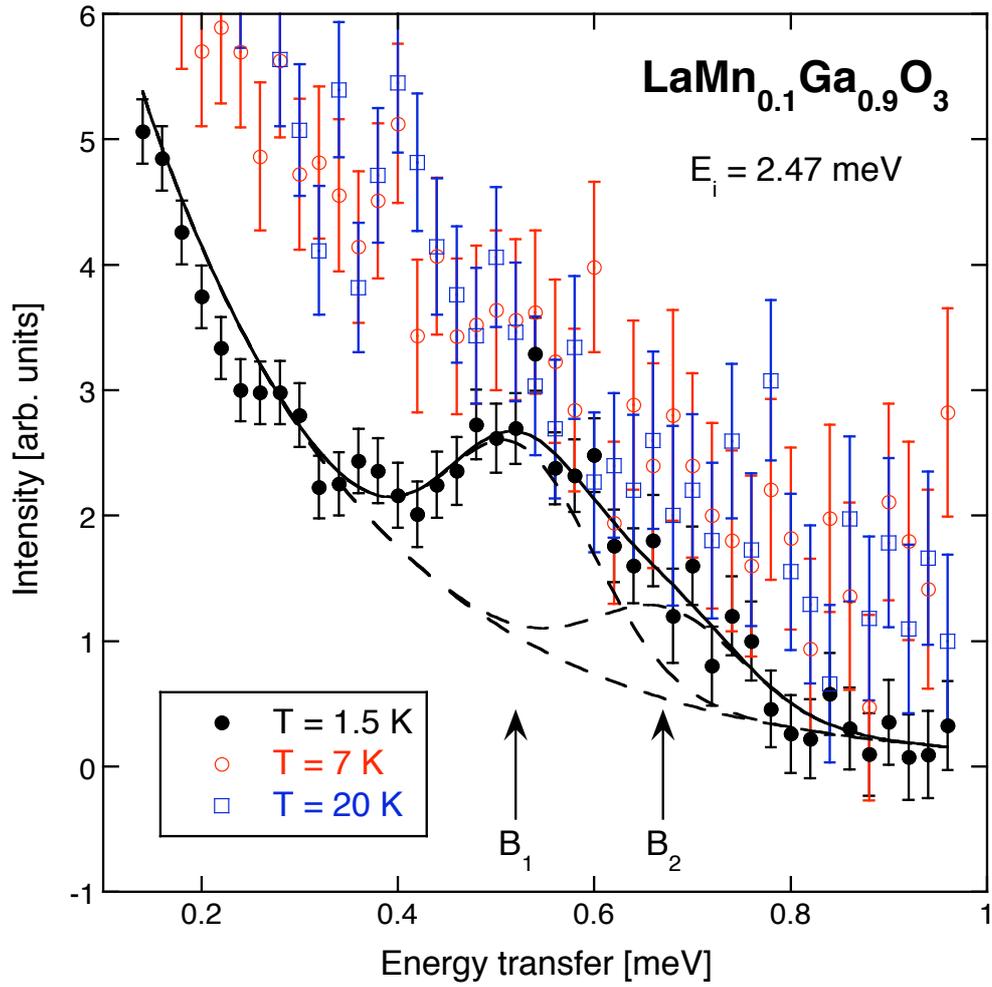

FIG. 4. (Color online) Energy spectra of neutrons scattered from $LaMn_{0.1}Ga_{0.9}O_3$ at different temperatures. The incoming neutron energy was 2.47 meV. The lines are the result of a least-squares fitting procedure as explained in the text.

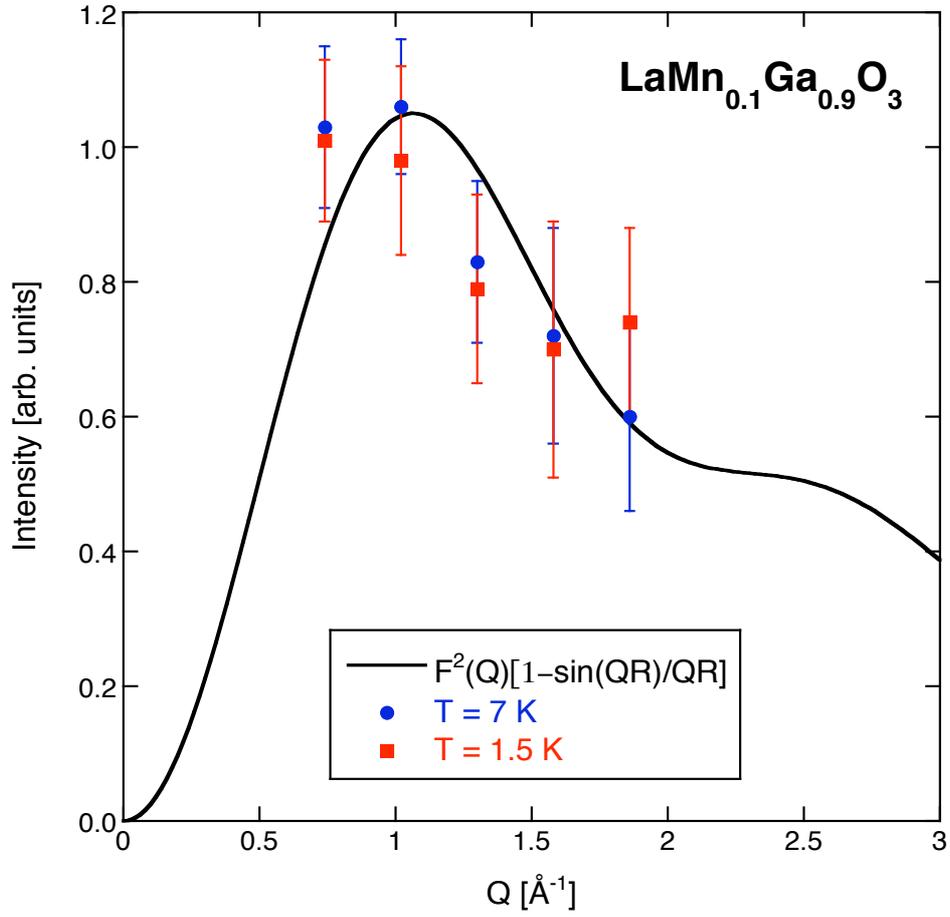

FIG. 5. (Color online) Q dependence of the neutron cross-section associated with transitions of ferromagnetically coupled $Mn^{3+}$ dimers in $LaMn_{0.1}Ga_{0.9}O_3$. The line corresponds to Eq. (4).